\documentclass[conference, 10pt]{IEEEtran}

\usepackage{cite}
\usepackage{url}
\usepackage{graphicx}
\usepackage{color}
\usepackage{placeins}
\usepackage{float}
\usepackage{tabularx,colortbl}
\usepackage{ifthen}
\usepackage{hft-thesis}

\RequirePackage[caption=false,font=footnotesize]{subfig}
\makeatletter

\newcounter{author}
\renewcommand{\author}[2][]{
   \stepcounter{author}
   \@namedef{author@\theauthor}{#2}
   \@namedef{authorlabel@\theauthor}{#1}
}

\newcounter{address}
\newcommand{\address}[2][]{
   \stepcounter{address}
   \@namedef{address@\theaddress}{#2}
   \@namedef{addresslabel@\theaddress}{#1}
}

\newcommand{\alsep}{and}

\def\newmaketitle{\par%
  \begingroup%
  \normalfont%
  \def\thefootnote{}
  \def\footnotemark{}
  \let\@makefnmark\relax
  \footnotesize
  \footnotesep 0.7\baselineskip
  \normalsize%
  \twocolumn[\thenewmaketitle\@IEEEaftertitletext]%
  \if@IEEEusingpubid
     \enlargethispage{-\@IEEEpubidpullup}%
  \fi
  \endgroup
  \setcounter{footnote}{0}\let\maketitle\relax\let\@maketitle\relax
  \gdef\@thanks{}%
  \let\thanks\relax}

\def\thenewmaketitle{
  \newpage
  \begin{center}%
    \vskip0.2em{\Huge\@IEEEcompsoconly{\sffamily}\@IEEEcompsocconfonly{\normalfont\normalsize\vskip 2\@IEEEnormalsizeunitybaselineskip
   \bfseries\large}\@title\par}\vskip1.0em\par%
    \vspace{1ex}
    \newcounter{c@author}
    \newcounter{c@tmp}
    \ifthenelse{\value{author}=2}{%
      \newcommand{\liand}{ and }}{%
      \newcommand{\liand}{, and }}
    \ifthenelse{\value{address}<2}{%
      \@nameuse{author@1}%
      \stepcounter{c@author}%
      \whiledo{\value{c@author}<\value{author}}{%
        \setcounter{c@tmp}{\value{author}}%
        \addtocounter{c@tmp}{-\value{c@author}}%
        \ifthenelse{\value{c@tmp}=1}{%
          \renewcommand{\alsep}{\liand}}{\renewcommand{\alsep}{, }}%
        \stepcounter{c@author}\alsep \@nameuse{author@\thec@author}}\\%
    }
    {
      \@nameuse{author@1}${}^{(\ref{\@nameuse{authorlabel@1}})}$%
      \stepcounter{c@author}%
      \whiledo{\value{c@author}<\value{author}}{%
      \setcounter{c@tmp}{\value{author}}%
      \addtocounter{c@tmp}{-\value{c@author}}%
      \ifthenelse{\value{c@tmp}=1}{%
        \renewcommand{\alsep}{\liand}}{\renewcommand{\alsep}{, }}%
      \stepcounter{c@author}\alsep \@nameuse{author@\thec@author}%
        ${}^{(\ref{\@nameuse{authorlabel@\thec@author}})}$%
      }
    }
    \vspace{0.2ex}

    \ifthenelse{\value{address}>0}{%
      \ifthenelse{\value{address}=1}{
        {\@nameuse{address@1}}
      }
      {
        \newcounter{c@address}

        \begin{center}
        \whiledo{\value{c@address}<\value{address}}
        {
          \refstepcounter{c@address}
            ${}^{(\thec@address)}$\,%
              \label{\@nameuse{addresslabel@\thec@address}}%
              \@nameuse{address@\thec@address}\\ %
        }
        \end{center}
      } 
    }
    {
      \relax
    }
  \end{center}
}

\makeatother

\title{Multipath Exploitation in Highly Reflective Environments for Enhanced Microwave Imaging via Inverse Source Reconstruction}

\author[org1]{Quanfeng Wang}
\author[org2]{Mei Song Tong}
\author[org1]{Thomas F. Eibert}

\address[org1]{Department of Electrical Engineering, School of Computation, Information and Technology,\\
Technical University of Munich, Munich, Germany, quanfeng.wang@tum.de}
\address[org2]{{Department of Electronic Science and Technology, Tongji University, Shanghai, China}}

\AddToHook{shipout/foreground}{
  \put(0.5\paperwidth, -0.8cm){ 
    \makebox[0pt][c]{
      \begin{minipage}{\textwidth}
        \centering \scriptsize 
          This article has been accepted for publication in the 8th IEEE International Conference on Computational Electromagnetics (ICCEM). This is the author's version which has not been fully edited and content may change prior to final publication.
      \end{minipage}
    }
  }
}

\AddToHookNext{shipout/foreground}{
  \put(0.5\paperwidth, -\paperheight + 0.7cm){ 
    \makebox[0pt][c]{
      \begin{minipage}{\textwidth}
        \centering \scriptsize
          Copyright~\copyright~2026 IEEE. Personal use of this material is permitted. Permission from IEEE must be obtained for all other uses, in any current or future media, including\\reprinting/republishing this material for advertising or promotional purposes, creating new collective works, for resale or redistribution to servers or lists, or reuse of any copyrighted component of this work in other works by sending a request to pubs-permissions@ieee.org.
      \end{minipage}
    }
  }
}

\begin{document}

\newmaketitle

\begin{abstract}
Multipath effects significantly influence the quality of microwave imaging in highly reflective environments, while the physical measurement aperture size constrains resolution. It is shown that by exploiting multipath reflections, improved resolution can be achieved while maintaining acceptable artifact levels. Based on image theory, strong scattered fields from an ideal reflection plane can be represented by virtual image sources. Using a single-frequency inverse source solver, the spatially distributed original and image sources are reconstructed and separated, which allows separate application of the imaging algorithm for both of them. The coherent combination of both sets of sources together with appropriate phase correction results in an effective aperture expansion that yields superior resolution. Furthermore, this separation strategy significantly mitigates interference artifacts. Simulation results, supported by theoretical analysis and comparison with a ray-tracing enhanced back-projection algorithm are presented to verify the effectiveness of the proposed approach.
\end{abstract}

\section{Introduction}
In microwave imaging, multipath effects arising either from the targets of interest (TOIs) themselves~\cite{wang20253da} or from highly reflective environments~\cite{leigsnering2014multipatha}, can severely degrade the quality of the obtained images, e.g., in the form of image distortions or ghosting artifacts. Efforts have been devoted to multipath suppression to achieve high-contrast and ghost-free images. For instance, by examining the distinct phase histories of the targets from different viewing angles, ghosts can be distinguished from true TOIs, and subsequently suppressed~\cite{garren2002sar}. A similar concept is adopted in~\cite{obuchon2004drift}, where spatial drifting is considered as a criterion for ghost identification. However, these methods rely on a large number of radar aspect angles for accurate suppression. Additionally, based on the observation that true TOIs should be independent of the scanning configuration, an image fusion technique for measurements from multiple locations~\cite{ahmad2008multilocation} and an array rotation technique~\cite{guo2018multipath} have been proposed. While effective, these techniques are constrained by the substantial measurement overhead required at different locations or angles.

Compared with multipath suppression which discards potentially useful information contained in multipath contributions, utilizing this information for imaging enhancement is an attractive choice, known as multipath exploitation~\cite{leigsnering2014multipatha}. On the one hand, in certain scenarios where line-of-sight (LOS) propagation is blocked, multipath reflections become the sole contributions available for imaging~\cite{setlur2014multipath}. On the other hand, the energy and information contained in multipath signals can be leveraged to improve the signal-to-clutter ratio and image resolution. In~\cite{setlur2011multipath}, by using geometrical optics analysis combined with prior knowledge of the environment, the locations of associated ghosts can be determined and mapped back to the true TOIs, yielding high-contrast images. However, these works~\cite{setlur2014multipath,setlur2011multipath} are limited to 2-D scenarios and the process of determining ghost locations becomes significantly more intricate when extended to 3-D cases with complex environments.

A ray-tracing (RT) enhanced back-projection algorithm (RT-BPA) is introduced in~\cite{na2025polarizationaware}, where a general RT engine employing shooting and bouncing rays (SBR) methods, is used to generate possible LOS and reflected paths up to various orders. These paths are subsequently utilized to augment the BPA for imaging purposes, where it has been demonstrated that incorporating contributions from higher-order reflections can improve imaging resolution. However, as the reflection order increases, the ray density becomes increasingly sparse, making it progressively more difficult for the SBR method to identify valid paths. Additionally, while the LOS and reflected paths are separated, the total co-polarized received fields are used indiscriminately for BPA-based adjoint imaging across different reflection orders, which inevitably introduces additional artifacts when combining higher-order reflections for the final image.

In this work, leveraging image theory, an inverse source solver is employed to separate virtual image sources generated by reflections of different orders in highly reflectively environments. Subsequently, the imaging algorithm is applied individually to each image source. It has been demonstrated that such operator inversion techniques can yield significant benefits in imaging applications~\cite{wang2024TAP,saurer2025solution}. Finally, the image sources are coherently combined with the original LOS source with corresponding phase corrections, thereby achieving enhanced imaging with improved resolution and suppressed artifacts. In order to demonstrate these concepts, radiation problems are considered throughout this work. 

In Section~\ref{section2}, the inverse source reconstruction (ISR) and imaging algorithms are introduced. Numerical results and corresponding analyses are presented in Section~\ref{section3}, while some conclusions are drawn in Section~\ref{section4}.

\section{Imaging Algorithm}
\label{section2}

The basic imaging configuration is illustrated in Fig.~\ref{fig:Image_Theory}, where dipoles serving as TOIs are positioned adjacent to a sufficiently large perfectly electrically conducting (PEC) plate. A sufficient number of field samples are collected via planar measurements. According to image theory, the received electric field is equivalent to the superposition of the fields $\veg{E}_{0}$ and $\veg{E}_{1}$ radiated by the original sources and the image sources in free space, respectively. Therefore, the observed signal at $\veg{r}_{m}$ is expressed as (a time-dependence $\e^ {\jm 2\uppi ft}$ is suppressed) 
\begin{equation}
  \veg{E}\left({\veg{r}_m}\right)= \iiint_{V}  \vecop{G}\left(\veg{r}_m,\veg{r'}\right)\cdot \left[\veg{J}_{ 0}\left(\veg{r}'_{0}\right)+\veg{J}_{1}\left(\veg{r}'_{1}\right)\right]\,\dd v'\,  \label{eq:meas}
\end{equation}
where $\vecop{G}$ denotes the dyadic free-space Green's function, and $\veg{J}_{0}$ and $\veg{J}_{1}$ are equivalent current densities of  the original and the image sources, respectively. For more complex configurations involving higher-order reflections, additional equivalent sources $\veg{J}_{i}$ can be introduced. With the capability of the inverse source solver to reconstruct and separate spatially distributed sources~\cite{wang2024TAP,eibert2015electromagnetic}, the equivalent sources can be reconstructed separately in the form of plane-wave spectra (PWS) $\tilde{\veg{J}}_{i}(\veg{k},\veg{r}'_{i})$ for each discrete frequency $\mbox{$f=1,\dots,F$}$ based on
\begin{align}
   \veg{E}\left(\veg{r}_m\right)=\frac{-\jm}{4\uppi}  \oiint \left[ \sum_{\veg{r}'_{i}} T_{L}\left(\veg{k},\veg{r}_{m}-\veg{r}'_{i}\right)\,{\tilde{\veg{J}_{i}}(\veg{k},\veg{r}'_{i})}\right] \dd^2 \hat{k}  \,,\label{eq:FMM}
\end{align}
where $T_L$ is the well-known fast multipole translation operator of order $L$~\cite{Chew2001} and $\veg{k}=k\hat{ k}$ denotes the wave vector. With the obtained separate PWS of the original and image sources, a hierarchical disaggregation based imaging algorithm can be applied individually according to~\cite{wang2024TAP}
\begin{equation}
  {\mathring{\veg{J}}_{i}\left(k_{f},\veg{r}\right)}=\oiint \mathcal{F}\left(\hat{k}\cdot\hat{k}^{\mathrm{(c)}}\right) {\tilde{\veg{J}}_{\rm{i}}(\veg{k}_f,\veg{r}'_{i})}\e^{\jm \veg{k}_f \cdot (\veg{r}'_{0}-\veg{r}'_{i})} \e^{-\jm \veg{k}_f \cdot \veg{r}} \, \dd^2 \hat{k}\,,\label{eq:imageGen}
\end{equation}
where $\mathcal{F}\left(\hat{k}\cdot\hat{k}^{\mathrm{(c)}}\right)$ serves as an angular spectral windowing filter, managing the trade-off between resolution and artifact suppression~\cite{wang2024TAP}. In~\eqref{eq:imageGen}, the additional phase term $\e^{\jm \veg{k}_f \cdot (\veg{r}'_{0}-\veg{r}'_{i})} $ translates the virtual source to the true source position via the Fourier shift theorem.
Finally, the multi-frequency image generation is expressed as
\begin{align}
   {J}_{p}(\veg{r})=\sum_{f}\sum_{i}&\psi_{p}\,{\mathring{{J}}_{i\,,p}\left(k_{f},\veg{r}\right)}\,, \label{eq:sumF}
\end{align}
where $p\in\left\{x,y,z\right\}$ denotes the three vector components in Cartesian coordinates. According to image theory, the phase correction term $\psi_{p}$ for the image sources is set to -1 for parallel polarization and 1 for vertical polarization. In contrast to the RT-BPA~\cite{na2025polarizationaware}, where $\psi_{p}$ must be identified for every individual propagating path, the proposed method requires considering these relations only once for each image source.

A primary advantage of ISR based imaging compared to RT-BPA is the distinct separation of distributed sources, which significantly mitigates imaging artifacts. The BPA considering only co-polarized components of a radiation problem is expressed as 
\begin{equation}
  {s}^{\text{BPA}}_{p}(\veg{r})= \iint_{\text{rx}} {E}_{p}\left({\veg{r}_m}\right) \e^{\jm k \left|\veg{r}_{m}-\veg{r}\right|} \dd \veg{r}_{m}\,.\label{eq:BPA}
\end{equation}
Substituting the co-polarized scalar form of~\eqref{eq:meas} into~\eqref{eq:BPA} and interchanging the order of integration yields
\begin{align}
  {s}^{\text{BPA}}_{p}(\veg{r})&=\iiint_{V}\left[{J}_{0,\,p}\left(\veg{r}'_{0}\right)+{J}_{1,\,p}\left(\veg{r}'_{1}\right)\right] \notag\\ &\quad \quad\quad\quad\iint_{\text{rx}}  \op{G}_{pp}\left(\veg{r}_m,\veg{r'}\right) \e^{\jm k \left|\veg{r}_{m}-\veg{r}\right|} \, \dd \veg{r}_{m}\,\dd v'\,.
\end{align}
This reveals that the reconstruction quality relies on the second integral term ideally approaching the delta distribution \mbox{$\delta(\veg{r}-\veg{r}')$}, which is much harder to achieve with BPA than with explicit inversion. Consequently, the indiscriminate use of total co-polarized received fields ${E_p}\left({\veg{r}_m}\right)$ in RT-BPA inevitably introduces interference artifacts from ${J}_{1,\,p}$ to ${J}_{0,\,p}$, and vice versa. With the inclusion of more higher-order image sources, the resulting image becomes highly corrupted.
\section{Numerical Results and Analysis}
\begin{figure}[t]
	\centering
	\includegraphics[scale=0.65]{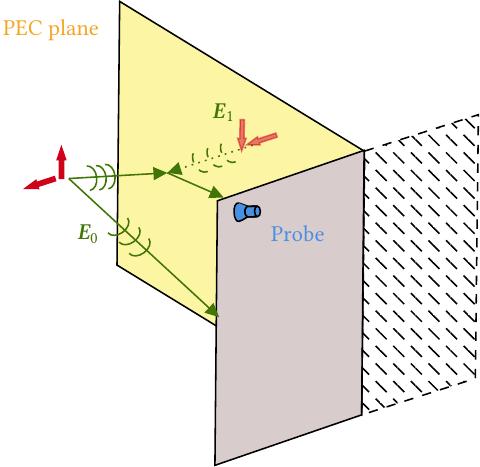}
	\caption{Imaging configuration of the sources and their corresponding image sources generated by reflection from the PEC plane. Multipath exploitation enables the expansion of the physical measurement aperture.}
	\label{fig:Image_Theory}
\end{figure}%
\label{section3}
\begin{figure}[t]
	\centering
	\includegraphics[scale=0.7]{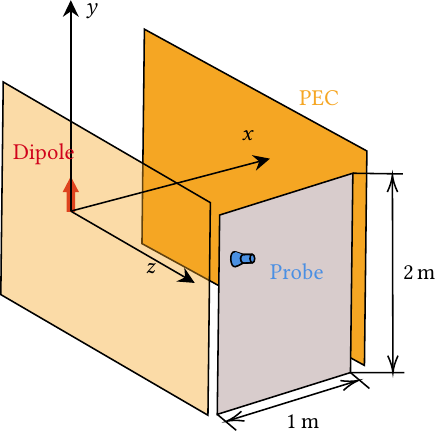}
	\caption{Illustration of the simulation setup with a Hertzian dipole placed between two PEC planes.}
	\label{fig:Simu1}
\end{figure}%
\begin{figure}[t]
  \centering
  \subfloat[]{\includegraphics[scale=0.59]{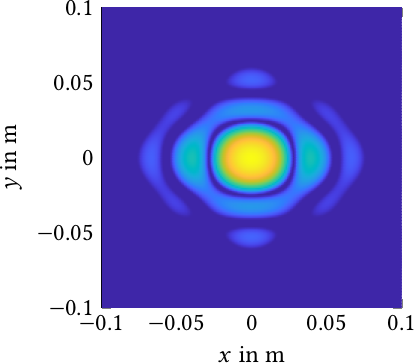}}%
  \hfill
  \subfloat[]{\includegraphics[scale=0.59]{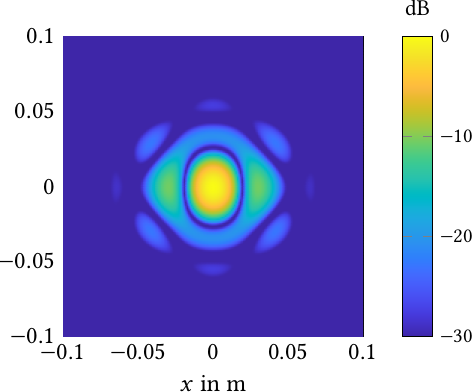}}%
  \\ 
  \vspace{-5mm}
  \subfloat[]{\includegraphics[scale=0.59]{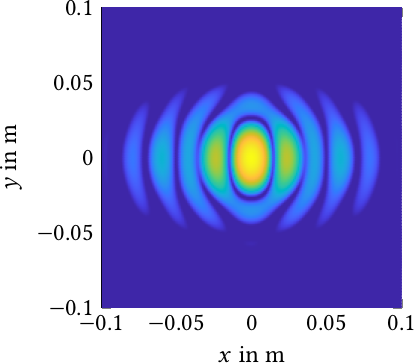}}%
   \hfill
  \subfloat[]{\includegraphics[scale=0.59]{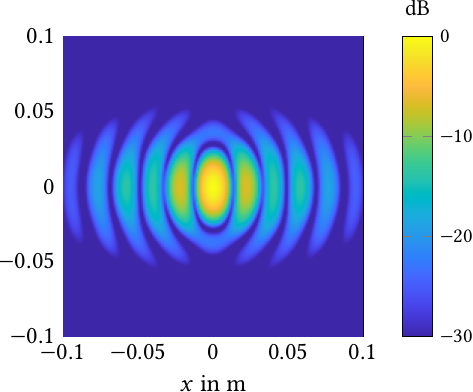}}%
  \caption{Imaging results of the single dipole based on the reconstructed original source (a), and incorporating virtual image sources corresponding to reflections up to first order~(b), second order~(c) and third order~(d), respectively. }
  \label{fig:plate2}
\end{figure}

\begin{figure}[t]
  \centering
  \subfloat[]{\includegraphics[scale=0.59]{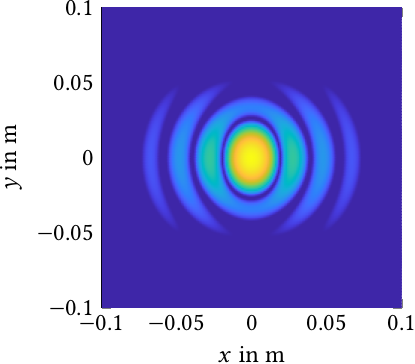}}%
  \hfill
  \subfloat[]{\includegraphics[scale=0.59]{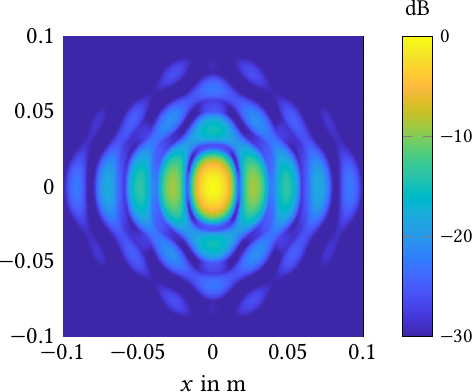}}%
  \caption{Reference results for Fig.~\ref{fig:plate2}(b).
  (a)~Simulation in free space with an enlarged measurement plane corresponding to the first-order reflection.
  (b)~RT-BPA imaging result considering LOS and first-order reflection contributions.}
  \label{fig:Plate2_ref}
\end{figure}
Full-wave simulations were performed using the commercial software FEKO~\cite{EMSS2025}. The first setup modeled after~\cite{na2025polarizationaware} is illustrated in Fig.~\ref{fig:Simu1}, where a $y$-polarized dipole is located at the origin between two sufficiently large parallel PEC plates. Eleven frequencies ranging from $\SI{8}{\giga\hertz}$ to $\SI{10}{\giga\hertz}$ with a step size of $\SI{200}{\mega\hertz}$ were selected. The electric fields consisting of the $x$- and $y$-components were acquired over a $\SI{1}{\meter}\times \SI{2}{\meter}$ planar surface at $z=\SI{1}{\meter}$, utilizing $20\,000$ uniformly distributed sampling points. According to symmetry, the coherent combination of image sources is equivalent to expanding the physical measurement aperture, thereby enhancing the theoretical resolution, as illustrated in Fig.~\ref{fig:Image_Theory}. 

The imaging results for the original source and combined image sources up to specific orders are shown in Fig.~\ref{fig:plate2}. It is observed that the inclusion of higher-order image sources yields a significant improvement in resolution along the $x$-axis. While the physical aperture is twice as large in the $y$-dimension, the two first-order image sources effectively result in a threefold expansion of the aperture in the $x$-dimension, leading to the resolution improvement observed in Fig.~\ref{fig:plate2}(b). This is validated by the free-space reference case in Fig.~\ref{fig:Plate2_ref}(a), which confirms a comparable resolution capability. Furthermore, a comparison with the RT-BPA result in Fig.~\ref{fig:Plate2_ref}(b) highlights the superior artifact suppression of the ISR method.

\begin{figure}[t]
	\centering
	\includegraphics[scale=0.7]{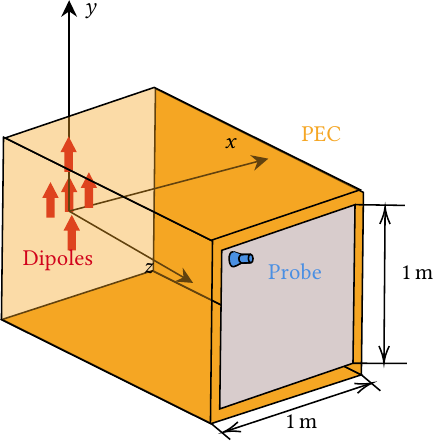}
	\caption{Illustration of the simulation setup with five Hertzian dipoles placed around the origin enclosing by four PEC planes.}
	\label{fig:Simu2}
\end{figure}%

\begin{figure}[t]
  \centering
  \subfloat[]{\includegraphics[scale=0.59]{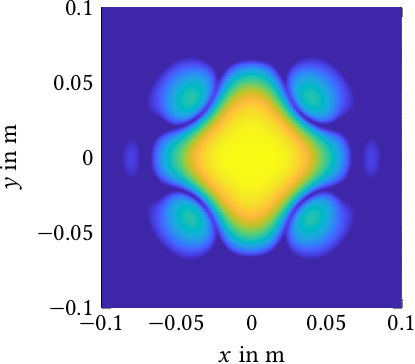}}%
  \hfill
  \subfloat[]{\includegraphics[scale=0.59]{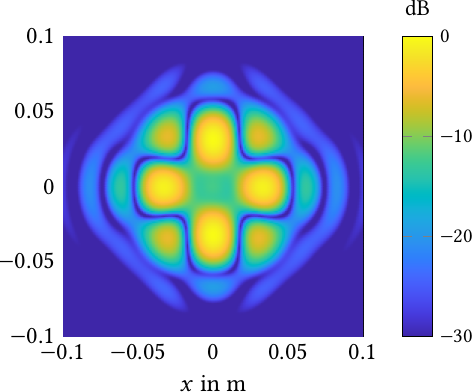}}%
  \caption{Imaging results of the five dipoles surrounded by four PEC plates. 
  (a)~Result based on the reconstructed original source. 
  (b)~Result incorporating virtual image sources corresponding to first- and second-order reflections.}
  \label{fig:plate4}
\end{figure}

\begin{figure}[t]
  \centering
  \subfloat[]{\includegraphics[scale=0.59]{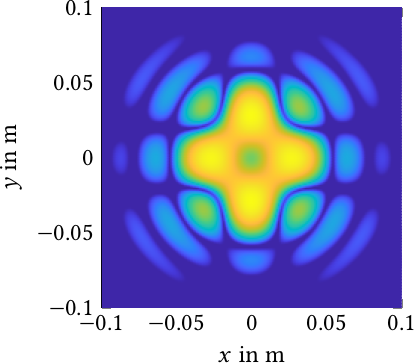}}%
   \hfill
  \subfloat[]{\includegraphics[scale=0.59]{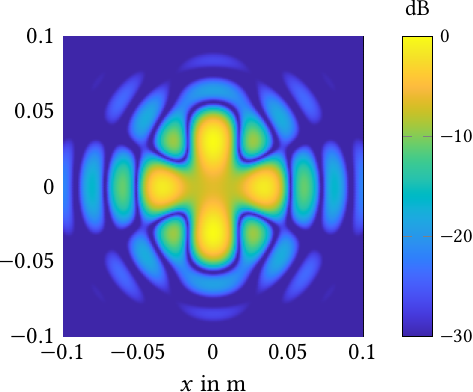}}%
  \caption{Reference results of Fig.~\ref{fig:plate4} obtained by removing the PEC plates and using different measurement plane sizes, (a) original size and (b) enlarged by a factor of three.}
  \label{fig:Plate4_ref}
\end{figure}

\begin{figure}[t]
  \centering
  \subfloat[]{\includegraphics[scale=0.59]{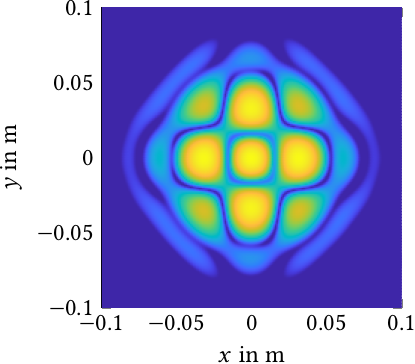}}%
  \hfill
  \subfloat[]{\includegraphics[scale=0.59]{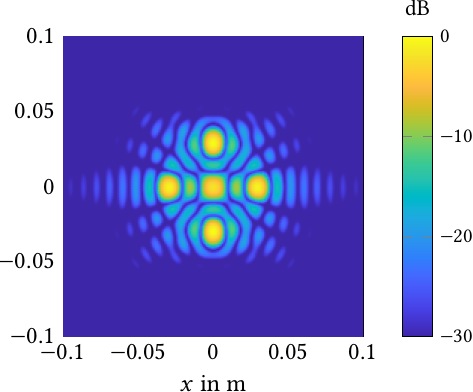}}%
  \caption{(a)~Imaging results of four dipoles surrounded by four PEC plates in the frequency band ranging from $\SI{8}{\giga\hertz}$ to $\SI{10}{\giga\hertz}$.
  (b)~Imaging results of the five dipoles at $\SI{20}{\giga\hertz}$.
  \label{fig:Plate4_reason}
  }
\end{figure}

The second setup comprises four PEC plates enclosing five $y$-polarized dipoles, spaced $\SI{0.03}{\meter}$ in the $x$- or $y$-direction. This distance is slightly smaller than the theoretical resolution~\cite{wang2024TAP}.
The measurement domain is a square plane containing $10\,000$ uniformly distributed sampling points. The imaging result for the original true sources is presented in Fig.~\ref{fig:plate4}(a), while a free-space reference is provided in Fig.~\ref{fig:Plate4_ref}(a). Both results fail to effectively resolve the five dipoles due to their dense arrangement. Especially, in Fig.~\ref{fig:plate4}(a), the dipoles are merged into a single indistinguishable region, whereas the free-space case shows only marginal improvement. This is because capturing all the image sources in such a highly reflective environment is practically infeasible. Consequently, the ISR process reconstructs only the lower-order image sources, treating the remainder as noise. This results in a worse signal-to-noise ratio, which inevitably degrades the imaging resolution~\cite{wang2024TAP}. Nonetheless, by incorporating image sources up to the second order, which is equivalent to a threefold increase of the aperture in both the $x$- and $y$-dimension, the dipoles become clearly distinguishable. The corresponding free-space reference is shown in Fig.~\ref{fig:Plate4_ref}(b).

It is observed in Fig.~\ref{fig:plate4}(b) and Fig.~\ref{fig:Plate4_ref}(b) that the dipole located at the origin is virtually invisible. This phenomenon arises from the specific dipole arrangement, where the four surrounding dipoles introduce sidelobes with opposite phase at the origin due to the point spread effect. Coincidentally, the magnitude of these superimposed sidelobes is comparable to that of the central dipole itself, thus, the resulting destructive interference suppresses the central dipole and makes it undetectable. To verify this, the central dipole was removed from the simulation and the resulting image of the four remaining dipoles is shown in Fig.~\ref{fig:Plate4_reason}(a), where a ghost dipole that does not actually exist emerges. This artifact is specific to this geometric configuration and resolution. Increasing the frequency to $\SI{20}{\giga\hertz}$ eliminates the phenomenon, as demonstrated in Fig.~\ref{fig:Plate4_reason}(b).

\section{Conclusion}
\label{section4}
An ISR based multipath exploitation method for microwave imaging in highly reflective environments has been presented. Leveraging image theory, the inverse source solver is applied to reconstruct both the true sources and the virtual image sources generated by reflections. The method allows for the distinct separation and individual imaging of spatially distributed sources. Furthermore, through appropriate phase correction, the original and image sources were coherently combined to enhance imaging performance. The results demonstrate significant resolution improvement, while artifacts are effectively suppressed compared to the RT-BPA method.

\section*{ACKNOWLEDGEMENT}
Funded by the European Union. Views and opinions expressed are however those of the author(s) only and do not necessarily reflect those of the European Union or European Innovation Council and SMEs Executive Agency (EISMEA). Neither the European Union nor the granting authority can be held responsible for them. Grant Agreement No: 101099491.



%

\bibliographystyle{IEEEtran}
\bibliography{Literature.bib}

\end{document}